\documentclass[epj]{svjour}
\usepackage{graphics}
\usepackage[T1]{fontenc}
\usepackage[latin9]{inputenc}
\usepackage{amsbsy}
\usepackage{graphicx}
\usepackage{amssymb}
\usepackage{amsmath}

\DeclareMathOperator{\dist}{dist}
\DeclareMathOperator{\Real}{Re}

\newcommand{\be}{\begin{equation}}
\newcommand{\ee}{\end{equation}}
\newcommand{\bel}[1]{\begin{equation}\label{#1}}
\newcommand{\bea}{\begin{eqnarray}}
\newcommand{\eea}{\end{eqnarray}}
\newcommand{\bef}{\begin{figure}}
\newcommand{\enf}{\end{figure}}
\newcommand{\ba}{\begin{array}}
\newcommand{\ball}{\begin{array}{ll}}
\newcommand{\bacl}{\begin{array}{cl}}
\newcommand{\bacll}{\begin{array}{cll}}
\newcommand{\bal}{\begin{array}{l}}
\newcommand{\bac}{\begin{array}{c}}
\newcommand{\ea}{\end{array}}

\newcommand{\R}{\mathbb{R}}

\renewcommand{\i}{\mathrm i}

\begin{document}
\title{Speed of complex network synchronization}
\author{Carsten Grabow\inst{1} \and Stefan Grosskinsky\inst{2} \and Marc Timme\inst{1,3,4}
}                     
\offprints{}          
\institute{Network Dynamics Group, Max Planck Institute for Dynamics and Self-Organization,
37073 G\"ottingen, Germany \and Centre for Complexity Science and Mathematics Institute, University of Warwick, Coventry CV4
7AL, UK \and Bernstein Center for Computational
Neuroscience (BCCN) G\"ottingen, 37073 G\"ottingen, Germany \and
Faculty of Physics, University G\"ottingen, 37077 G\"ottingen, Germany}
\date{Received: date / Revised version: date}
%

\abstract{Synchrony is one of the most common dynamical states
emerging on networks. The speed of convergence towards synchrony
provides a fundamental collective time scale for synchronizing
systems. Here we study the asymptotic synchronization times for
directed networks with topologies ranging from completely
ordered, grid-like, to completely disordered, random,
including intermediate, partially disordered topologies. We
extend the approach of Master Stability Functions to quantify
synchronization times. We find that the synchronization times
strongly and systematically depend on the network topology. In
particular, at fixed in-degree, stronger topological randomness
induces faster synchronization, whereas at fixed path length,
synchronization is slowest for intermediate randomness in the
small-world regime. Randomly rewiring real-world neural, social
and transport networks confirms this picture.
} 

\maketitle

\section{Introduction and overview}

We live in a world where everything is connected: we are surrounded by global networks of communication, transportation, trade, social relations and media. In addition, due to the enormous progress in technology, we are now able to decipher natural networks of high complexity. For example, we get to know more details about the brain or the human genome day by day. But understanding the function of these networks requires a two-sided approach. Firstly, we need to know the basic structures of these networks at both microscopic and macroscopic level. Secondly, we have to map the rules which govern the dynamic interactions. Although all these networks seem to be different at first sight, there are similarities such as abstract patterns or simple organizing principles \cite{Arenas:2008p1192}. It is not only the individual unit that matters but also the architecture of the connections. 
To uncover these rules, theoretical studies first focus on systems consisting of simple units such as oscillators and of simple structures: all-to-all coupled units, lattices, or mean field models.
Although exact results on synchronization in networks with a general
structure have been obtained recently \cite{BarahonaPecora,Earl,Timme02},
it is still not well understood how the structure of a complex network
affects dynamical features of synchronization.

But a question which dates back to the sixties changed this approach, namely: `what is the
probability that any two people, selected arbitrarily from a large population,
such as that of the United States, will know each other?' \cite{Milgram}.
A more interesting formulation, however, takes into account that, while two persons
may not know each other directly, they may share one or more mutual
acquaintances such that any two people are connected through a chain of acquaintances. This concept, known as 'six degrees of separation', refers to the idea that everyone is on average approximately six steps away from any other person on Earth. In 1998, this idea was converted into a simple model of small-world networks \cite{WattsStrogatz98}. The crucial point is that this model interpolates between totally regular and totally random topologies.

Starting with a ring where units only communicate with their direct neighbours, ring connections between neighbours are cut and connected to randomly chosen nodes somewhere else in the ring. An important quantity which characterizes this network's evolving architecture is the 'average path length' and formalizes the intuitive idea of degrees of separation. Due to the rewiring, the path length between any pair of units drastically decreases. Simultaneously this architecture still exhibits high local clustering, defined as the probability that two nodes linked to a common node will also be linked to each other. These two properties were suggested to be particularly supportive of synchronization. Indeed, several detailed studies support this view by showing that at fixed coupling strength small-world networks tend to already synchronize at lower connectivity than many other classes of networks \cite{BarahonaPecora,WattsStrogatz98}. 

Synchronization is one of the most frequently observed collective dynamics in many physical
and biological systems \cite{Arenas:2008p1192,SynchBook,Sync}. Synchronization might be both advantageous
and desired, for instance in secure communication \cite{Kanter:2002p2846},
or detrimental and undesired, as during tremor in patients with Parkinson
disease or during epileptic seizures \cite{Maistrenko:2004p1453,Netoff:2004p241}. \\
Therefore, a broad area of research has emerged \cite{Strogatz01,Nishikawa:2003p656,Pecora:1998p265}, 
determining under which conditions on the interaction strengths and topologies coupled units actually synchronize and when they do not. These results suggest some key properties about the topological influence on the network synchronizability, i.e. the capability of a network to synchronize at all. They do not tell much about the speed of synchronization given that a network synchronizes in principle.

For any real system, however, it equally matters how fast the units synchronize
or whether the network interactions fail to coordinate the units'
dynamics on time scales relevant to the system's function (or dysfunction), cf. \cite{Zumdieck:1990p2850,Zillmer:2007p2849,Jahnke:2008p2847,Zillmer:2009p2851}. The applications range from consensus dynamics of distributed decision-making problems for interacting groups of agents \cite{Olfati2005} to neuroscience questions of how fast the visual processing or olfactory discrimination could be \cite{uchida,thorpe}. Yet this question is far from being understood and currently under active investigation \cite{Grabow10,Timme04,Timme06,Timme:2006p292,Qi:2008p1201,Qi:2008p1198}.
In particular it is largely unknown how fast small worlds synchronize which leads us to our main question addressed here: what is the typical time scale for synchronization, i.e. how fast can oscillators coordinate their dynamics if they are not directly interconnected
but interact on large networks of regular, random or small-world topology? 

We address this question by computer simulations as well as analytical predictions. All results are derived for the simplest
of all regular states, the synchronous periodic state, in which all
oscillators exhibit identical dynamics. However, also other settings are imaginable: cluster states in which two or more groups
of synchronized oscillators exist \cite{Ernst:1995:1570,Ernst:1998:2150} or systems with inhomogeneities in the dynamical and topological parameters \cite{Denker:2004p2853} can be treated similarly.

We study the effect of topology on the synchronization time of directed networks which exhibit different dynamics: Kuramoto phase oscillators coupled via phase differences, higher-dimensional periodic R\"ossler systems coupled diffusively as well as neural circuits with inhibitory delayed pulse-coupling.
Synchronization time is a measure of how quickly the network synchronizes after being perturbed from a synchronized state. So far it has been studied analytically for fully random networks only \cite{Timme06}. 
Firstly, comparing network ensembles with a fixed number of edges, it is shown that those in the small-world regime synchronize faster than regular networks but slower than random networks. This is expected intuitively -- the characteristic path length is monotonically decreasing while rewiring -- and in accordance with the result for synchronizability \cite{BarahonaPecora,WattsStrogatz98}. Hence, we fix the average characteristic path length and again investigate the dependence of synchronization time on the network's topology. We find that -- for a fixed average characteristic path length -- networks in the small-world regime again synchronize slower than random networks, but this time even slower than regular networks: we see a non-monotonic dependence on the topological randomness. First results have been reported in \cite{Grabow10} and here we further systematically investigate these studies and extend them to real-world networks. We compare network ensembles with fixed topological quantities like the betweenness centrality as well as generic ensembles for Kuramoto, R\"ossler and pulse-coupled oscillators. Moreover, we make analytical predictions of the synchronization times for periodic R\"ossler systems and observe remarkable similarities between the synchronization times for the Kuramoto and pulse-coupled oscillators. 

This article is organized as follows. In Section \ref{synctime} we first introduce the concept of synchronization time, the central quantity of our studies. In Section \ref{top} we explain the underlying network structure and introduce the small-world model adapted to directed networks. Section \ref{models} outlines the different types of considered dynamics. Here we derive analytical predictions for the synchronization times and extend the master stability function formalism \cite{Pecora:1998p265} to determine the synchronization speed. In Section \ref{ensembles} we compare the analytical predictions for the synchronization times and the results obtained by extensive computer simulations for network ensembles with fixed in-degree, with fixed average path length and with fixed betweenness centrality, followed by an analysis of generic network ensembles. In Section \ref{real} the study of synchronization times for real-world networks -- rewiring them towards fully random networks -- confirms our theoretical results. We close in Section \ref{conc} with a summary and a discussion of further work.

\section{Synchronization time}\label{synctime}

\begin{figure*}
\includegraphics[width=1.0\textwidth]{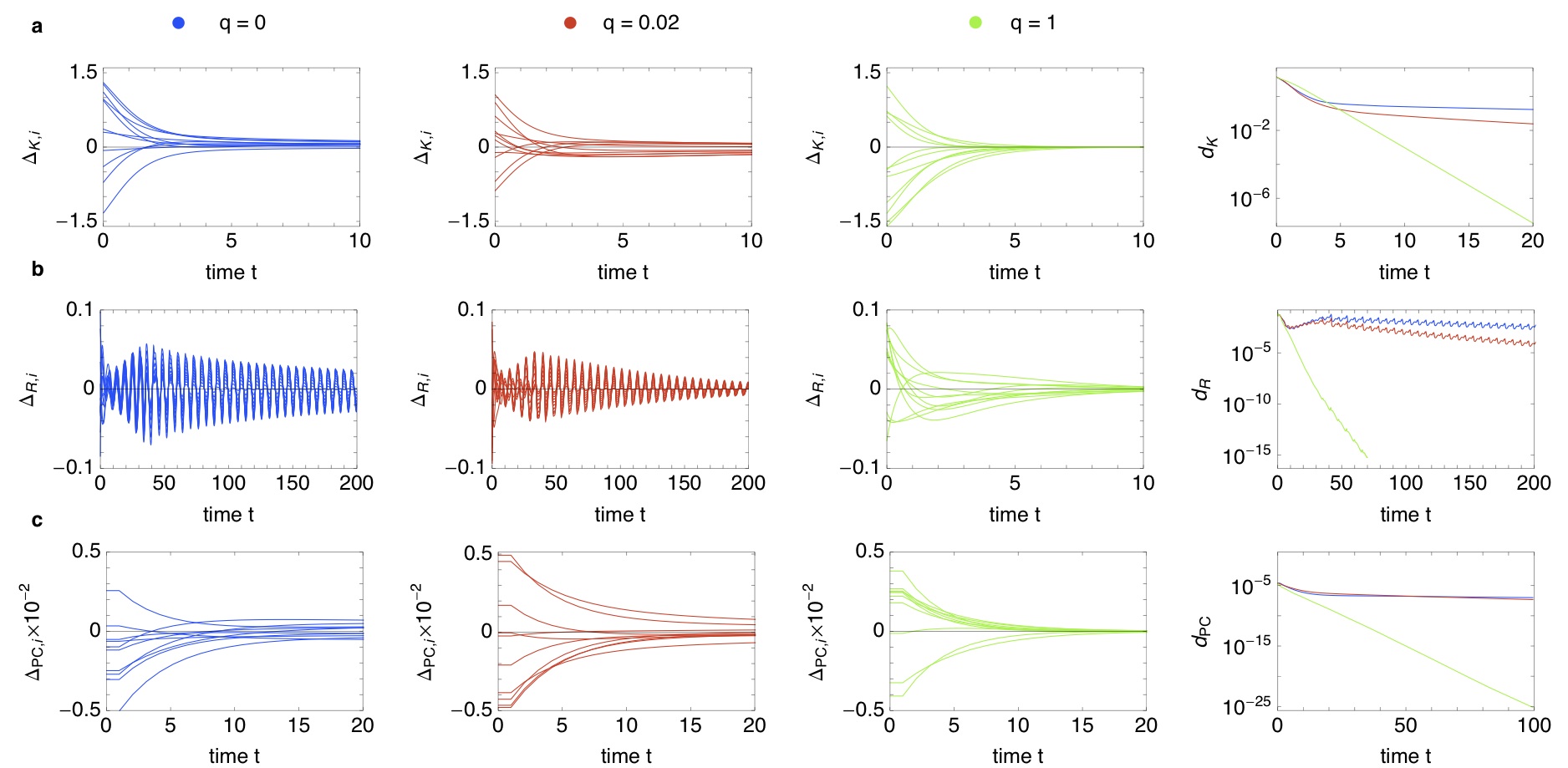}
\caption{(Color online) Time scales of synchronization of oscillator networks for topological randomness $q\in\{0, 0.02, 1\}$ (in-degree fixed at $k=20$). \textbf{top row a}: Kuramoto oscillators ($\sigma=1$); \textbf{b}: diffusively coupled periodic R\"ossler oscillators ($a=0.2$, $c=5.7$, $b=1.7$, $\sigma=2$); \textbf{c}: pulse-coupled oscillators ($I=1.01$, $\gamma=1$, $\sigma=-0.2$, $\Delta=0.1$). See equations~(\ref{eq:kuradeltas}) --~(\ref{eq:PCdeltas}) for the definitions of the variable differences. Plotted in the right-hand column are the logarithmized decaying distances (see equations (\ref{eq:distancedecay}), (\ref{eq:kuradistance}), (\ref{eq:roesslerdist}) and (\ref{eq:PCdistance}))}.
\label{fig:alldeltas}     
\end{figure*}

We assume identical oscillators which means we get complete synchronization in the end. The equation of motion for the uncoupled oscillators looks like
\begin{equation}
\frac{d\vec{x}_i}{dt} =\vec{F} (\vec{x}_i) \ ,
\label{eq:freeoscillators}
\end{equation}
where the $m$-dimensional vector $\vec{x}_{i} = \{x_{i,1}, ..., x_{i,m}\}$ refers to the components of each oscillator $i \in \{1,...,N\}$ and $\vec{F}: \R^{m}\mapsto\R^{m}$ defines the dynamics. 

We describe the connection of $N$ oscillators in a network by a coupling matrix $J$ that basically consists of zero and non-zero elements to specify which oscillators are coupled to which other ones. The matrix elements $J_{ij}$ are thus given by 

\begin{equation}
J_{ij}=\left\{ \begin{array}{ll}
\sigma/k_{i} & \mbox{if $j$ is connected to $i\neq j$}\\
0 & \mbox{otherwise}\, \end{array}\right. \ , \label{eq:Jmatrix}
\end{equation}
where $\sigma$ is a global coupling constant and $1/k_{i}$ a normalization factor that guarantees a homogeneous total input
\begin{equation}
\sum_{j=1}^{N}J_{ij}=k_{i} \frac{\sigma}{k_{i}}=\sigma 
\label{eq:couplingnorm}
\end{equation}
such that no specific oscillator receives distinguished couplings.

Directly related to the coupling matrix $J$ is the graph Laplacian $\Lambda$ defined as
\begin{equation}
\Lambda_{ij}=J_{ij}(1-\delta_{ij}) -  \sigma \delta_{ij} \ , \label{eq:laplacian}
\end{equation}
where $\delta_{ij}$ is the Kronecker-delta. 
Considering directed networks its eigenvalues $\lambda_{i}$ are complex and ordered as $0 = \Real \lambda_{1}\ge\Real \lambda_{2}\ge\ldots\ge\Real \lambda_{N}$. The number of zero eigenvalues of the Laplacian matrix is equal to the number of strongly-connected components (SCCs) of the network\footnote{A strongly-connected component is the maximal subnetwork such that for each given
pair of nodes $(i,j)$ there is a directed path from $i$ to $j$ and from $j$ to $i$.}. It is trivial then to conclude that if $\Real \lambda_{2}=0$, the network is split in more than one SCC. Then, from a dynamical point of view, it is impossible for the network to achieve a complete synchronized state, which is only possible for subnetworks with internal coherence. We are considering networks with one SCC only throughout this article, which means we have always $\Real \lambda_{2}> 0$. Note that for the pulse-coupled system in Section \ref{PC} the eigenvalues are ordered according to their absolute values.

We describe the dynamics of the interaction with a function $\vec{H}$ that is a vector function of dimension $m$ of the dynamical variables of two connected oscillators. Each oscillator has the same interaction function. For example, $\vec{H}$ for the R\"ossler oscillators is a $3 \times 3$ matrix that only picks out the $x$-component to couple to the other oscillators. The coupled equations of motion become
\begin{equation}
\frac{d\vec{x}_i}{dt} =\vec{F}(\vec{x}_i) + \sum_{j=1}^{N} J_{ij} \vec{H}(\vec{x}_{i},\vec{x}_{j}) \ ,
\label{eq:coupledoscillators}
\end{equation}
where $J_{ij}>0$ acts on each oscillator as a whole. Furthermore, note again that we only consider identical oscillators which means that the global coupling $\sigma$ is the same for each oscillator as well as the interaction function $\vec{H}$. Since we want to examine the case of identical synchronization, the equations of motion become the same for all oscillators when the system is synchronized. 
In the synchronous state all oscillators' variables are equal to the same dynamical variable: 
\begin{equation}
\vec{x}_1(t) = \vec{x}_2(t) = \ldots = \vec{x}_N(t) = \vec{s}(t) \ ,
\label{eq:synchstate}
\end{equation} 
where $\vec{s}(t)$ is a solution of (\ref{eq:freeoscillators}) as long as $\vec{H}(\vec{s}(t),\vec{s}(t))=0$, which is the case for Kuramoto and R\"ossler oscillators. The subspace defined by the constraint of setting all oscillator vectors to the same, synchronous vector is called the synchronization manifold. We assume stability of this state which means that small arbitrary perturbations to each $\vec{x}_j$ die out in the long time limit.

In addition to these dynamical systems with continuous-time coupling we introduce pulse-coupled systems as well in Section \ref{PC}.

We consider directed regular, small-world and random networks which are characterized by increasing rewiring, the topological randomness $q$. By tuning this parameter we interpolate between regular ring networks ($q = 0$), small worlds (low $q \ll1$) and fully random networks ($q = 1$) which is explained in detail in Section \ref{top}.

First simulations for three different kinds of oscillators (see Fig.~\ref{fig:alldeltas}) show
that synchronization becomes an exponential process after some short
transients for all fractions
$q\in [0,1]$ of randomness. 
Thus the distance 
\begin{equation}
d(t)=\max_{i,j}\dist(\vec{x}_{i}(t),\vec{x}_{j}(t))\label{eq:distancedefinition}
\end{equation}
from the synchronous state decays as
\begin{equation}
d(t)\sim\exp(-t/\tau)\label{eq:distancedecay}
\end{equation}
in the long time limit, where $\dist(\vec{x},\vec{x}')$
is a function measuring the distance between the two appropriate phase variables $\vec{x}$ and $\vec{x}'$, taking into account the periodic boundary conditions. The characteristic time scale $\tau$ in (\ref{eq:distancedecay}) is what we call the synchronization time in the following. Note that there exist systems as well where the transient until the exponential decay is not negligible \cite{Zumdieck:1990p2850,Jahnke:2008p2847,Tonjes:2010io}.

As one can see in Fig.~\ref{fig:alldeltas} this decay is similar for Kuramoto, R\"ossler and pulse-coupled oscillators. It depicts the differences of the phase variables (which we defined in detail in Section \ref{models}) of ten randomly chosen oscillators to the corresponding means denoted by $\left[ \ . \ \right]$:
\begin{align}
\Delta_{\textrm{K},i}(t)&=\Theta_{i}(t)-\left[ \Theta_{j}(t) \right]_{j}, \label{eq:kuradeltas}\\
\Delta_{\textrm{R},i}(t)&=x_{i}(t)-\left[ x_{j}(t) \right]_{j},  \label{eq:Rdeltas} \\ 
\Delta_{\textrm{PC},i}(t)&=\tilde \phi_i(t)-\left[ \tilde \phi_j(t) \right]_{j},  \label{eq:PCdeltas}
\end{align}
with
\be
\tilde \phi_i(t)=\left\{ \begin{array}{ll}
\phi_i(t) & \textrm{ if $\phi_i(t)\leq0.5$} \ ,\\
\phi_i(t)-1 & \textrm{ if $\phi_i(t)>0.5$}
\end{array} \right. 
\ee
'K' stands for Kuramoto, 'R' for R\"ossler and 'PC' for pulse-coupled oscillators and these abbreviations will be kept throughout this paper.

In contrast to the continuous-time dynamics of the Kuramoto and R\"ossler oscillators, for the pulse-coupled oscillators the phases are measured at discrete 'spiking' times of a reference oscillator.
For the $3$-dimensional R\"ossler oscillators only the $x$-coordinates are shown here. 
The actual dynamical variables for all systems will be introduced in Section \ref{models}.

\section{Network topology}\label{top}

\begin{figure}
\begin{centering}
\includegraphics[width=80mm]{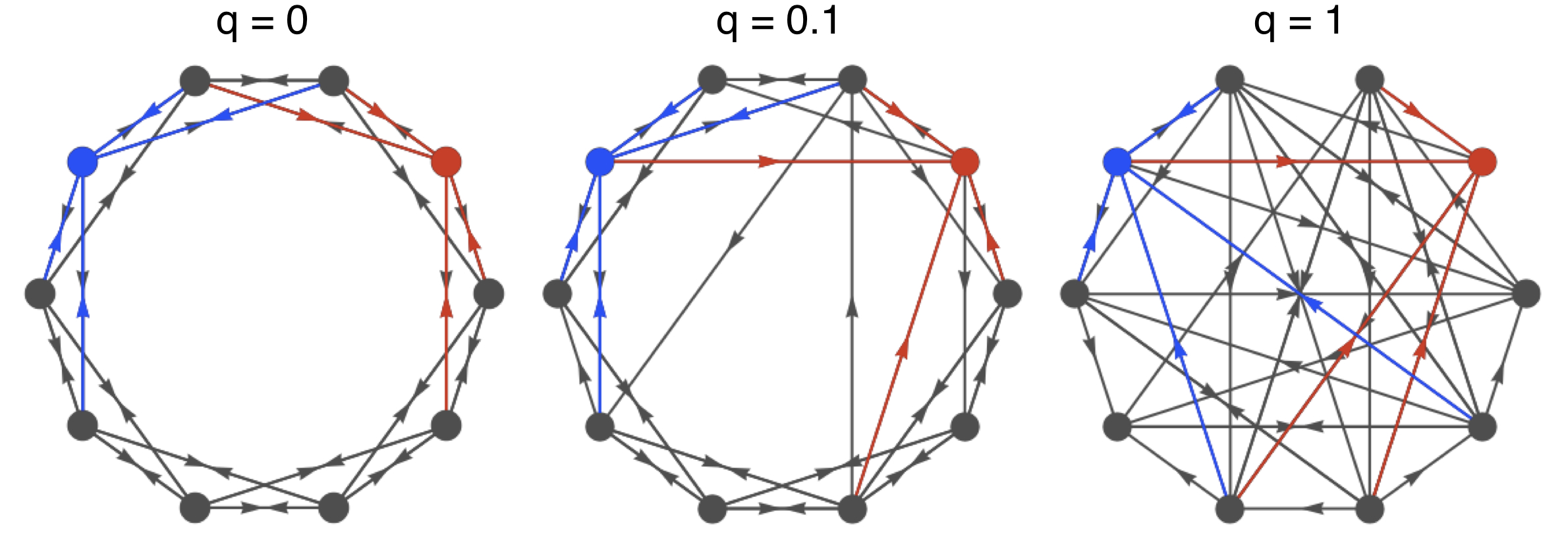}
\par\end{centering}
\caption{(Color online) Rewiring directed networks, the in-degree $k$ for each node stays fixed: this can be observed here for two reference nodes' incoming edges depicted in blue and red. (Here: $N=10$, $k=4$, $q\in\{0, 0.1, 1\}$).
\label{fig:rewiring} }
\end{figure}

We adapt the standard small-world model of Watts and Strogatz \cite{WattsStrogatz98} to directed
networks \cite{clustdef}. 
We start with regular ring networks where each unit $i$ receives directed links from its $k_{i}/2$ nearest neighbors on both sides ($k$ is chosen to be even). Here the in-degree $k_{i}=k \ \forall i$ is the same for all units. We randomly cut each outgoing edge with probability $q$ and rewire it to a node chosen uniformly at random from the whole network (avoiding double edges and self-loops). We do, however, allow the edge to be rewired back to its original position. 
An important observation here is that as $q$ varies the in-degree of each node (and with it the average in-degree of the network) is still $k$ (see Fig.~\ref{fig:rewiring}). This is due to the fact we only rewire outgoing egdes. 
Furthermore the entries $J_{ij}\geq 0$ of the coupling matrix are multiplied by a global coupling constant $\sigma$ and normalized to guarantee that each oscillator $i$ is getting the same input as has already been mentioned in Section \ref{synctime}. The matrix elements $J_{ij}$ are therefore $J_{ij}=\sigma/k_{i}$ if there is a connection from $j$ to $i \neq j$, $J_{ij}=0$ if there is no connection and $J_{ii}=0$ for the diagonal elements (\ref{eq:Jmatrix}).

To analyze the purely topological impact on the synchronization times, we study the network
dynamics in its simplest setting: we consider strongly-connected networks with fixed in-degree $k$ and homogeneous total input coupling strengths (encoded in the coupling matrix $J$ (\ref{eq:Jmatrix})) such that full synchrony is achieved from sufficiently close initial conditions for all coupling strengths $\sigma$ \cite{Timme:2006p292}.

The directed small-world networks behave as in the original Watts-Strogatz model. 
The small-world regime ($q$ small) is characterized by a large clustering coefficient\footnote{$C(q,k)$ denotes the actual divided by the possible number of directed triangles containing a given node $i$, averaged over all $i$.} $\left\langle C(q,k)\right\rangle $ and a small average path length\footnote{$L(q,k)$ denotes the length of the shortest directed path between a given
pair of nodes $(i,j)$, averaged over all $(i,j)$.} $\left\langle L(q,k)\right\rangle$.
Here $\left\langle \ . \ \right\rangle $ denotes averaging over network
realizations at given $q$ and $k$. 

To quantitatively fix the small
world regime we take
\begin{equation}
\frac{\left\langle L(q,k)\right\rangle }{L(0,k)}<0.5\quad\mbox{and}\quad\frac{\left\langle C(q,k)\right\rangle }{C(0,k)}>0.85\label{eq:LCdefinitions}
\end{equation}
throughout this study. The results below are not sensitive to a change
of these values.

As the topological randomness $q$ is changed from $0$ to $1$ the network interpolates between regular and random topologies. 
This structural change induces changes in the corresponding graph Laplacian's spectrum and thus has a direct influence on the synchronization speed as is explained in detail in Section \ref{models}.

\section{Oscillator dynamics on networks}\label{models}

We consider various oscillator types, intrinsic dynamics and coupling schemes: phase oscillators coupled via phase differences, neural circuits with inhibitory delayed pulse-coupling and higher-dimensional periodic systems coupled diffusively.
In the three following subsections we introduce the theory of these different types of oscillators and add remarks on the simulations, the chosen initial conditions and the numerical measurement of the synchronization time.

\subsection{Kuramoto oscillators} \label{Kura}

Consider $N$ Kuramoto oscillators \cite{Acebron:2005p293} that interact
on a directed network. Here the dynamical variable of each oscillator is $\vec{x}_{i}:=\theta_{i}\in\mathbb{S}^{1}=2\pi \mathbb{R}/\mathbb{N}$, i.e. a one-dimensional phase, with its interaction function $\vec{H} (\theta_{i},\theta_{j}):=\sin(\theta_{j}-\theta_{i})$. Therefore the dynamics of phases $\theta_{i}(t)$ of oscillators $i$ with time $t$ satisfy 
\begin{equation}
\frac{d\theta_{i}}{dt}=\omega+ \sum_{j}J_{ij}\sin(\theta_{j}-\theta_{i})\hspace{0.5cm}\,\mbox{for}\, i\in\{1,...,N\}\ ,\label{eq:Kuramoto}\end{equation}

\noindent where $\omega$ is the natural frequency of the oscillators.
The fully synchronous state defined in (\ref{eq:synchstate}) here takes the form \bel{eq:synchstatekura}
\theta_{i}(t)\equiv\theta_{j}(t)=:\theta(t)
\end{equation} 
As the synchronous periodic orbit analyzed is isolated in state space, the relaxation time continuously changes with possible inhomogeneities, so the qualitative results obtained below are generic and also hold in the presence of small heterogeneities, cf.~\cite{Denker:2004p2853}.

Furthermore, starting from random initial phases in the range $[0,\pi]$ the synchronization dynamics shows a fast transient. After this fast initial evolution all phases are quite similar and the sine function in (\ref{eq:Kuramoto}) can be well approximated by its argument. Linearizing (\ref{eq:Kuramoto}) close to the synchronous state (\ref{eq:synchstatekura}) phase perturbations defined as
\begin{equation}
\delta_{\textrm{K},i}(t):=\theta_{i}(t)-\theta(t)
\end{equation}
evolve according to
\begin{equation}
\frac{d\delta_{\textrm{K},i}}{dt}= \sum_{j}\Lambda_{ij}\delta_{\textrm{K},j}(t)\hspace{0.5cm}\,\mbox{for}\, i\in\{1,...,N\}.\label{linearmodel}
\end{equation}
Here the stability matrix coincides with the graph Laplacian defined in (\ref{eq:laplacian}).

Close to every invariant trajectory the eigenvalue $\lambda_{2}$
of the stability matrix $\Lambda$ that is second largest in real
part dominates the asymptotic decay in the long time limit  
\begin{equation}
d_{\textrm{K}}(t)\sim\exp(-t/\tau_{\textrm{K}}) \ .
\label{eq:kuradistance}
\end{equation}
The distance $d_{\textrm{K}}(t)$ is given by (\ref{eq:distancedefinition})
where $\dist(\theta,\theta')$ for Kuramoto oscillators is the circular distance between the two phases $\theta$ and $\theta'$
on $\mathbb{S}^{1}$.

$\lambda_{2}$ here determines the asymptotic synchronization time which is given by
\bel{eq:kurasynctime}
\tau_\textrm{K}=-\frac{1}{\Real \lambda_{2}} .
\ee
This feature was recently shown to hold also more generally for network systems where the stability matrix is not necessarily proportional to the graph Laplacian \cite{Arenas:2008p1192,Timme04,Almendral:2007p1211}.

\subsection{R\"ossler oscillators} \label{Roessler}

Consider a network of R\"ossler oscillators, both in the chaotic and in the periodic regime. 
Each elementary oscillator is described now by three variables $\{{x}(t),{y}(t),{z}(t)\}$.
The collective dynamics of $N$ coupled, identical R\"ossler oscillators ($i \in \{1,2,...,N\}$) is governed by the
equations 
\begin{align}
\dot {x}_i=&-{y}_i-{z}_i + \sum_{j=1}^{N}J_{ij}({x}_j-{x}_i), \nonumber \\ 
\dot {y}_i=& {x}_i+a{y}_i, \nonumber  \\
\dot {z}_i=& b+{z}_i({x}_i-c),
\label{eq:roessler}
\end{align}
where $a$, $b$ and $c$ 
are fixed parameters. 

To study the R\"ossler system in the periodic regime we set the parameters to $a = 0.2$, $b = 1.7$, $c = 5.7$. Analogously setting the parameters to $a = 0.2$, $b = 0.2$, $c = 5.7$ the chaotic attractor is gained.

The evolution of perturbations is characterized by measuring the Euclidean distances
\begin{equation}
d_{ij}(t)= \sqrt{({x}_i(t){-}{x}_j(t))^2 {+}({y}_i(t){-}{y}_j(t))^2{+}({z}_i(t){-}{z}_j(t))^2}
\label{eq:roesslereuclidean}
\end{equation}

between the states of all $N(N-1)/2$ possible pairs of oscillators $(i,j)$.
The asymptotic synchronization time is then determined via the decay of the maximal distance
\begin{equation}
d_{\textrm{R}}(t)=\max_{i,j} d_{ij}(t).
\label{eq:roesslerdist}
\end{equation} 

\begin{figure}
\begin{centering}
\includegraphics[width=80mm]{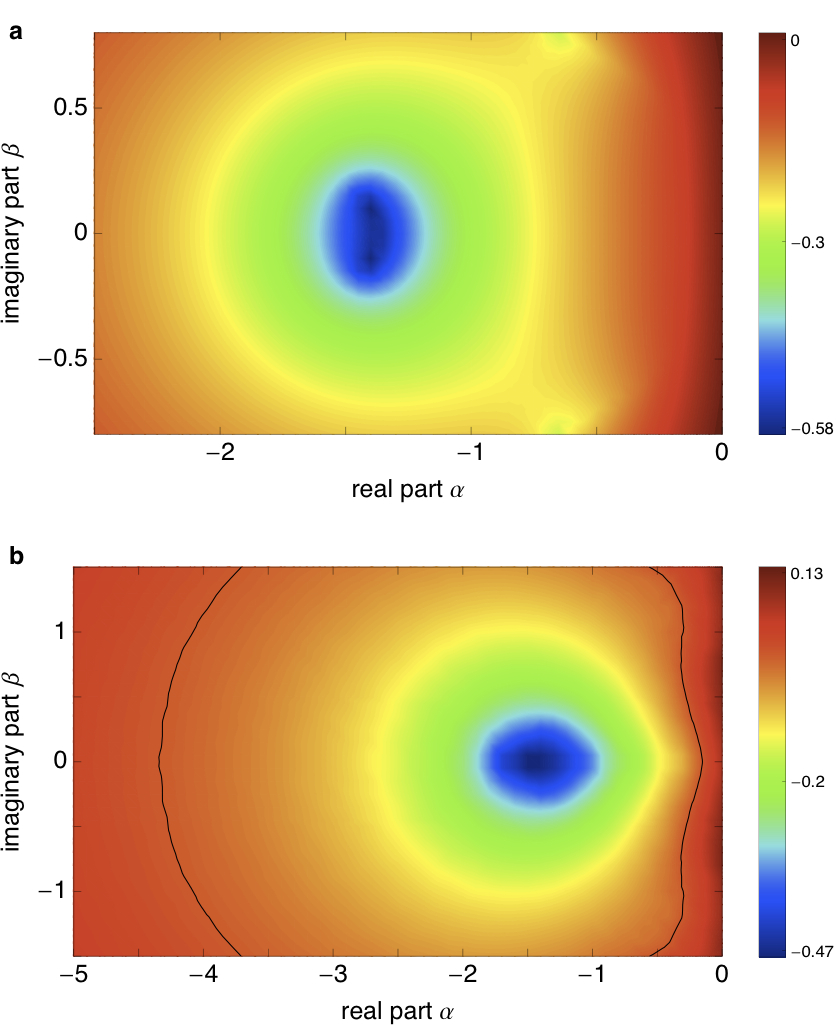} 
\par\end{centering}
\caption{(Color online)
Master stability functions (\ref{eq:hmax}) for the R\"ossler oscillators. \textbf{a}: periodic regime (parameters set to $a = 0.2$, $b = 1.7$, $c = 5.7$). \textbf{b}: chaotic regime (parameters set to $a = 0.2$, $b = 0.2$, $c = 5.7$; the black contour shows the MSF equal to zero, i.e. separates the stable from the unstable region).} 
\label{fig:MSF_pc}
\end{figure}

A general approach to determine the synchronization time for continuous systems described by (\ref{eq:coupledoscillators})-- alternative to the one taken for the Kuramoto oscillators, which does not work for the R\"ossler oscillators -- is to extend the master stability function (MSF) formalism introduced in \cite{Pecora:1998p265}. Note that this approach does not work for the pulse-coupled oscillators, where the phases are measured at discrete times. So far this formalism has only been used to determine the stability of networks of coupled oscillators \cite{Fink2000,Huang2009} and nearly all studies have focussed on symmetric undirected networks (see \cite{Hwang2005} for an exception). 

Defining infinitesimal perturbations to the synchronous state (\ref{eq:synchstate}) in the system described by Eq.~(\ref{eq:coupledoscillators}) as 
\bel{eq:roesslerperturb}
\boldsymbol{\delta}_{\textrm{R},i}=\vec{x}_{i}(t)-\vec{s}(t)
\ee
we get the variational equation
\begin{equation}
\frac{d\boldsymbol{\delta}_{\textrm{R},i}}{dt} = \vec{D} \vec{F} (\vec{s})\boldsymbol{\delta}_{\textrm{R},i} -  \sum_{j=1}^{N} \Lambda_{ij} \vec{D} \vec{H} (\vec{s},\vec{s})\boldsymbol{\delta}_{\textrm{R},i},
\label{eq:variational}
\end{equation} 

where the matrix $\Lambda$ is the graph Laplacian defined in (\ref{eq:laplacian}), $\vec{D} \vec{F}(\vec{s})$ and $\vec{D} \vec{H} (\vec{s},\vec{s})$ are the Jacobians evaluated along the trajectory $\vec{s}(t)$. 

For the above R\"ossler system with diffusive coupling via the $x$-coordinate the Jacobian matrices for this block are given by
\begin{equation}
\vec{D} \vec{F} (x,y,z)=\begin{pmatrix} 0 & -1 & -1 \\ 1 & a & 0 \\ z & 0 & x - c   \end{pmatrix}
\end{equation}
and
\begin{equation}
\vec{D} \vec{H} (x,y,z)=\begin{pmatrix} 1 & 0 & 0 \\ 0 & 0 & 0 \\ 0 & 0 & 0   \end{pmatrix}
\end{equation}

The transformation $\boldsymbol{\delta'}_{\textrm{R}}=O^{-1}\boldsymbol{\delta}_{\textrm{R}}$, where $O$ is a matrix whose columns are the set of the Laplacian's eigenvectors, diagonalizes the set of equations (\ref{eq:variational}) and hence leads to a set of decoupled blocks of the form
\begin{equation}
\frac{d\boldsymbol{\delta'}_{\textrm{R},i}}{dt} =\left[ \vec{D} \vec{F} (\vec{s}) -  \lambda_{i} \vec{D} \vec{H} (\vec{s},\vec{s})\right] \boldsymbol{\delta'}_{\textrm{R},i} ,
\label{eq:blocks}
\end{equation} 
with the $\lambda_{i}$ being the eigenvalues of the Laplacian matrix $\Lambda$. The above-given Jacobians evaluated in the synchronized state $\vec{s}(t)$ are the same for each block, hence the blocks only differ by the scalar multiplier $\lambda_{i}$.

Thus these blocks could be evaluated all at once by setting
\begin{equation}
\frac{d\boldsymbol{\delta'}_{\textrm{R},i}}{dt} =\left[ \vec{D} \vec{F} (\vec{s}) - (\alpha + \i \beta) \vec{D} \vec{H} (\vec{s},\vec{s})\right] \boldsymbol{\delta'}_{\textrm{R},i}
\label{eq:MSFalphabeta}
\end{equation}
in dependence on the complex coupling parameter $\alpha + \i \beta$. The imaginary part $\beta$ may be interpreted as a 'rotation' taking place between the several decaying eigenmodes of the system \cite{pecora97}.

\begin{figure}
\begin{centering}
\includegraphics[width=80mm]{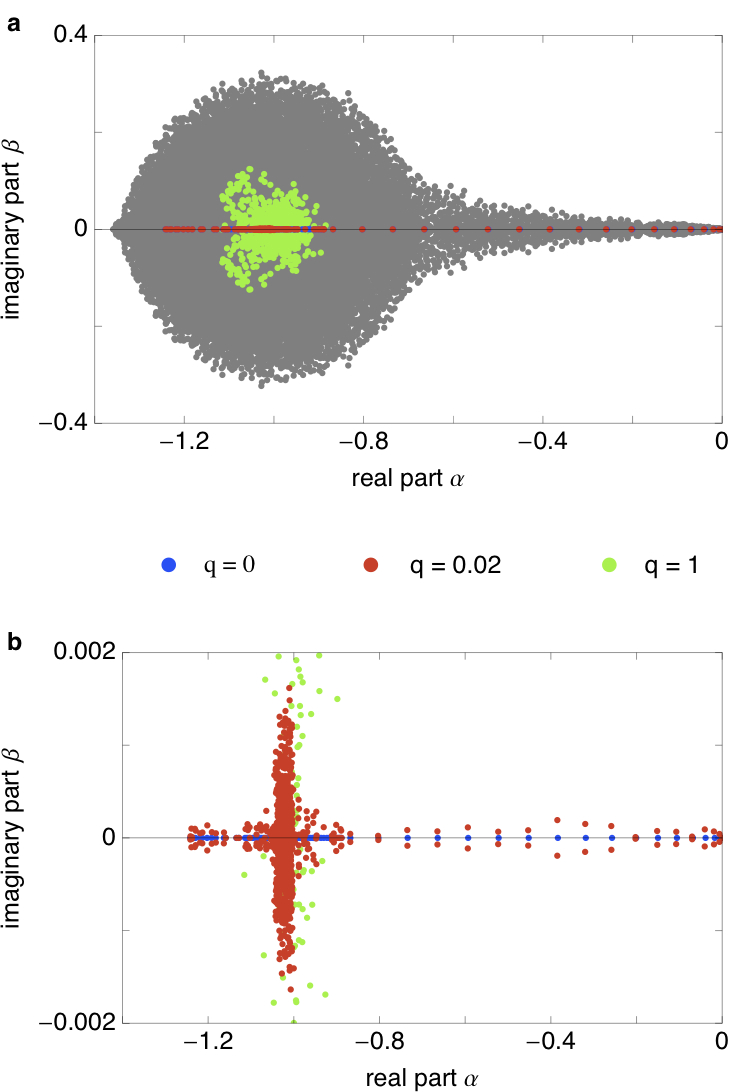} 
\par\end{centering}
\caption{(Color online) The graph Laplacian's eigenvalue distribution directly links to the system's stability and synchronization dynamics. \textbf{a}: in gray the whole complex spectrum of the graph Laplacians in the ranges $q \in [0,1]$ and $k \in [10,100]$, blue: eigenvalues (purely real due to the initial ring symmetry) for $q=0$, $k=50$, red: eigenvalues in the SW regime with $q=0.02$, $k=50$, green: eigenvalues for $q=1$, $k=50$. \textbf{b}: zoom to the real axis, eigenvalues for $k=50$ and $q\in\{0, 0.02, 1\}$ with colors as above.
\label{fig:eigenvalues} }
\end{figure}

The system actually synchronizes if 
\begin{equation}
h_{1,i} = \lim_{t \to \infty} \frac{1}{t}\log \frac{|\boldsymbol{\delta'}_{\textrm{R},i}(t)|}{|\boldsymbol{\delta'}_{\textrm{R},i}(0)|} < 0 \end{equation}
for all $i \in\{2,\ldots,N\}$.
Here $h_{1,i}$ is the largest Lyapunov exponent corresponding to the mode of eigenvalue $\lambda_{i}$ (see e.g. \cite{ott93}). 

To obtain the asymptotic synchronization time we extract the largest Lyapunov exponent $h_{1,i}$ with the minimal absolute value out of the $N-1$ maximal exponents, namely
\begin{equation}
h_{1,\max}=\max_{i \ge 2}h_{1,i}.
\label{eq:hmax}
\end{equation}
This is also called the master stability function and plotted in Fig.~\ref{fig:MSF_pc} for the periodic and chaotic R\"ossler oscillators. We calculate the largest Lyapunov exponent following the numerical procedure described in \cite{Huang2009}.
It is evident that the $\lambda_{1}=0$-mode is parallel to the synchronization manifold while all the other modes are transverse to it.
The synchronization time for the R\"ossler oscillators is then given by
\bel{eq:rsynctime}
\tau_\textrm{R} = -\frac{1}{h_{1, \max}}.
\ee

$h_{1, \max}$ dominates the decay towards the synchronized state, but note the nonlinear dependence on the eigenvalues of the Laplacian matrix (\ref{eq:blocks}). Only for the simple $1$-dimensional Kuramoto oscillators there is a direct relation, since here there is linear and unbounded coupling, i.e. the larger the global coupling the faster the synchronization speed.

In order to find a value for the global coupling parameter $\sigma$ - encoded in $\Lambda$ according to (\ref{eq:laplacian}) - that leads to synchronization, one calculates the whole spectrum of possible eigenvalues (see Fig.~\ref{fig:eigenvalues}) to guarantee that each one is located in the stable region, i.e. in the region where the MSF takes only negative values. Note that the MSF for uncoupled periodic R\"ossler oscillators is zero, while the MSF for uncoupled chaotic R\"ossler oscillators is positive. This means that the minimal global coupling constant needed to achieve synchronization is always larger for the chaotic R\"ossler oscillators than for the periodic ones.

\subsection{Pulse-coupled oscillators}\label{PC}

Moreover, we investigated the collective dynamics of pulse-coupled neural oscillators \cite{Jahnke:2008p2847,MIROLLO:1990p366}, which do not exactly fit the general description in Section \ref{synctime}.
In this case the dynamical oscillator variables are the membrane
potentials $V_{i}(t)$ and delayed discrete output pulses satisfying 
\bel{eq:membranedynamics}
\frac{dV_{i}}{dt}=I- \gamma V_{i}+  \sum_{j=1;\, j\neq i}^{N}\sum_{m\in\mathbb{Z}}J_{ij}\delta\left(t-\left(t_{j,m}+\Delta\right)\right).
\end{equation}
Here, each potential $V_{j}$ relaxes towards $I>1$ and is
reset to zero whenever it reaches a threshold at unity,
\begin{equation}
V_{j}(t^{-})=1\,\Rightarrow\, V_j (t):=0,\, t_{j,m}:=t,\,\mbox{and}\, m\mapsto m+1.\end{equation} 
At these times $t_{j,m}$ neuron $j$ sends a pulse that after a delay
$\Delta>0$ changes the potential of post-synaptic neurons $i$ in an
inhibitory (negative) manner according to (\ref{eq:membranedynamics}) with $\sigma<0$ in (\ref{eq:Jmatrix}).

Equivalent to these ordinary differential equations there is a simplified approach which represents the state of a one-dimensional oscillator not by its membrane potential, but by a phase
that encodes the time to the next spike in the absence of any interactions. The state of an individual oscillator $j$ is then represented by a phase-like
variable $\phi_{j}\in(-\infty,1]$ that increases uniformly in time,
\begin{equation}
d\phi_{j}/dt=1\,.\label{eq:phidot1}
\end{equation}
Upon crossing the firing threshold, $\phi_{j}(t^{-})=1$,
at time $t$ an oscillator is instantaneously reset to
zero, $\phi_{j}(t)=0$, and a pulse is sent. After
a delay time $\Delta$ this pulse is received by all oscillators $i$
connected to $j$ and induces an instantaneous
phase jump given by
\[
\phi_{i}((t+\Delta)^{+})=U^{-1}\left(U(\phi_{i}(t+\Delta)+ J_{ij}\right)\]
Here, the coupling strengths from $j$ to $i$ are taken to be purely inhibitory ($\sigma<0$ in (\ref{eq:Jmatrix})) and normalized according to (\ref{eq:couplingnorm}).
The rise function $U$, which mediates the interactions, can be derived from (\ref{eq:membranedynamics}) \cite{Timmechaos}, and turns out to be monotonic
increasing, $U'>0$, concave (down), $U''<0$, and represents the
subthreshold dynamics of individual oscillators. Note that the function $U$ need to be defined on the entire
range of accessible phase values. In particular, inhibitory coupling
can lead to negative phase values $\phi_{i}<0$. 

The synchronous state $\vec{s}(t)$ defined in (\ref{eq:synchstate}) here takes the form
\begin{equation}
\phi_{i}(t)=\phi_{0}(t)
\end{equation}
for all $i$, which is a self-consistent solution assuming that all neuronal oscillators fire at the same time. Here all oscillators display identical phases $\phi_{0}(t)$ on
a periodic orbit such that $\phi_{0}(t+T)=\phi_{0}(t)$ with the period 
\begin{equation}
T=\Delta+1-\alpha\label{eq:period}
\end{equation}
where 
\begin{equation}
\alpha=U^{-1}(U(\Delta)+ \sigma).\label{eq:alpha}
\end{equation}
Note that here in contrast to the Kuramoto oscillators the period is different from the one of a free oscillator \cite{Timme02}.

A perturbation 
\begin{equation}
\boldsymbol{\delta}_{\textrm{PC}}(0)=:\boldsymbol{\delta}_{\textrm{PC}}=(\delta_{\textrm{PC},1},\ldots,\delta_{\textrm{PC},N})
\end{equation}
to the phases is defined as 
\begin{equation}
\delta_{\textrm{PC,i}}=\phi_{i}(0)-\phi_{0}(0)\,.\label{eq:define_delta}
\end{equation} 

The initial condition for the phases of the pulse-coupled oscillators is a random perturbation $\boldsymbol{\delta}_{\textrm{PC}}$ from the globally synchronized state $\boldsymbol{\delta}_{\textrm{PC}}=\vec{0}$.
The perturbation's components $\delta_{\textrm{PC},i}$ are each drawn independently from a uniform distribution on $\left[-\delta,\delta\right]$. 
The condition $\delta<\frac{\Delta}{2}$ derived in \cite{Timme06} (recall that $\Delta$ is the delay time) ensures that the globally synchronized state is stable. This guarantees that all the neurons fire before any spikes are received. 

A sufficiently small perturbation $\boldsymbol{\delta}_{\textrm{PC}}$ asymptotically
converges exponentially with time to a constant vector. Subtracting the asymptotic phase shift,
\begin{equation}
\boldsymbol{\delta}'_{\textrm{PC}}(t):=\boldsymbol{\delta}_{\textrm{PC}}(t)-\lim_{s\rightarrow\infty}\boldsymbol{\delta}_{\textrm{PC}}(s),
\label{eq:phaseshift}
\end{equation}
the distance 
\begin{equation}
d_{\textrm{PC}}(nT):=\max_{i}|\delta'_{\textrm{PC},i}(nT)|
\label{eq:PCdistance}
\end{equation}
from the synchronous state ($\delta'_{\textrm{PC},i}\equiv0$) decays as
\begin{equation}
d_{\textrm{PC}}(nT)\sim \exp(-\frac{nT}{\tau_{\textrm{PC}}})
\label{eq:delta_tau}
\end{equation}
as $n\to\infty$, defining a synchronization time $\tau_{\textrm{PC}}$. 

To understand how the speed of synchronization depends on the dynamical
and network parameters, we analyze how perturbations $\boldsymbol{\delta}_{\textrm{PC}}$ to the synchronous
state evolve in time.
Following \cite{Timme02} we first define a nonlinear stroboscopic
map 
\bel{eq:strob}
\boldsymbol{\delta}_{\textrm{PC}}(nT)=\vec{G}(\boldsymbol{\delta}_{\textrm{PC}}\big((n-1)T)\big)
\ee
for the perturbations. Note that $\boldsymbol{\delta}_{\textrm{PC}}(T)=\boldsymbol{\delta}_{\textrm{PC}}(0)$ since no spikes are received before all the oscillators reach the phase threshold for the first time. Hence we first apply the map $G$ in the first period when spikes are received i.e. for $n \ge 2$.

Considering the first order approximation of this period-$T$ map one gets
a linear iterative map $A$ given by
\begin{equation}
\delta_{\textrm{PC},i} (nT)=\sum_{j=1}^N A_{ij} \delta_{\textrm{PC},j} \big( (n-1)\, T\big)\ ,\quad n \ge 2,
\label{eq:iterativemap}
\end{equation}
for the perturbations $\delta_{\textrm{PC},i} (nT)$ of spike times
close to the synchronous orbit of period $T=\ln\big( I/(I-1)\big)$. 

The matrix elements $A_{ij}$ are defined as
\begin{equation}
A_{ij}=\left\{ \begin{array}{ll}
p_{i,m}-p_{i,m-1} & \mbox{if $j$ is connected to $i\neq j$}\\
p_{i,0} & \mbox{if}\, j=i\\
0 & \mbox{otherwise}\, \end{array}\right.\label{eq:matrixelements}
\end{equation}
where the variables $p_{i,m}$ ($m \in \{1,\ldots,k_{i}\}$) encode phase jumps evoked by all pulses
up to the $m\textrm{th}$ one received \cite{Timme02}. Since the matrix elements
(\ref{eq:matrixelements}) are differences of these $p_{i,m\,},$
matrix elements $A_{ij}$ and $A_{ij'}$ with $j\neq j'$ have in
general different values depending on the order of incoming signals. 

This multi-operator problem \cite{Timme08} is induced by the structure of the network
together with the pulsed interactions, in particular, by the order of the components of $\boldsymbol{\delta}(0)$. For networks with homogeneous, global coupling different matrices $A$ can be identified by an appropriate
permutation of the oscillator indices. But in general this is
impossible. However, here we focus on the integrate-and-fire dynamics where the matrix $A$ becomes independent of the rank order of the perturbations \cite{Timme06}. Here $U$ takes the form
\bel{eq:risefunction}
U(\phi):=\frac{I}{\gamma}(1-e^{-\gamma \phi}).
\end{equation}
In order to obtain the matrix elements $A_{ij}$ we first calculate
\begin{equation}
U'(\phi)=I e^{-\gamma \phi}
\end{equation}
and
\begin{equation}
U^{-1}(y)=\frac{1}{\gamma} \ln{(1-\frac{y \gamma}{I})^{-1}}.
\end{equation}
Furthermore we calculate
\begin{equation}
U^{-1}(U(\Delta)+J_{ij})= \frac{1}{\gamma} \ln{(e^{-\gamma\Delta}-\frac{\gamma}{I}  J_{ij})^{-1}} 
\end{equation} 
and
\begin{equation}
U'(U^{-1}(U(\Delta)+ J_{ij}))=I e^{-\gamma \Delta}-\gamma  J_{ij}.
\end{equation} 
This leads to
\begin{align} 
p_{i,m}:=&\frac{U'(U^{-1}(U(\Delta)+ \sum_{l=1}^{m} J_{ij_{l}}))}{U'(U^{-1}(U(\Delta)+ J_{ij}))} \nonumber \\
=&\frac{I e^{-\gamma \Delta}-\gamma   \sum_{l=1}^{m} J_{ij_{l}}}{I e^{-\gamma \Delta}-\gamma  J_{ij}}
\label{eq:fractions}
\end{align}
where the sum $\sum_{l=1}^{m} J_{ij_{l}}$ with $m \in \{1,\ldots,k_{i}\}$ counts up to the $m\textrm{th}$ signal received by neuron $i$ during the considered period.

For homogeneous inhibitory coupling, $\sigma/k<0$ for each existing connection, the elements of the
stability matrix are given by
\begin{equation}
A_{ij}=\left\{ \begin{array}{ll}
a_{0}/k & \mbox{if $j$ is connected to $i\neq j$}\\
1-a_{0} & \mbox{if}\, j=i\\
0 & \mbox{otherwise}\, \end{array}\right.\label{eq:PCmatrixelements}
\end{equation}
with
\begin{equation}
a_{0}=\frac{\gamma \sigma}{I e^{- \gamma \Delta} +\gamma \sigma} \ .
\end{equation}
Note $A$ is a stochastic matrix and all diagonal entries satisfy $A_{ii}>0$. Hence the matrix is aperiodic which implies that the eigenvalue $a_1=1$ is the largest and is unique. 

We let $\textbf{v}_{i}$ for $i=1,2,\ldots,N$ be the eigenvectors of $A$ with corresponding eigenvalues $\left|a_1\right|>\left|a_2\right|\ge\ldots\ge\left|a_N\right|$. The eigenvector corresponding to the eigenvalue $a_1=1$ is $\textbf{v}_{1}=\left(1,1,\ldots,1\right)^\mathsf{T}$ since the row-sums of $A$ are equal to one. 
Recall that this means the distance vector $d_{\textrm{PC}}(n)$ does not tend to zero as $n\rightarrow\infty$, but instead to a uniform phase shift (\ref{eq:phaseshift})
\begin{equation}
\lim_{s\rightarrow\infty}\boldsymbol{\delta}_{\textrm{PC}}(s)=:\boldsymbol{\delta}_{\infty}
\end{equation}
which has all components equal, $\left(\boldsymbol{\delta}_{\infty}\right)_i=\delta_{\infty}$ for all $i$ (i.e. all the neurons are at the same phase and hence in a globally synchronized state). 
Furthermore, recall that the distance from the globally synchronized state is given by 
\begin{equation}
d_{\textrm{PC}}(nT):=\max_{i}|\delta'_{\textrm{PC},i}(nT)|
\end{equation} 
as defined in (\ref{eq:PCdistance}). Using the fact that $a_1=1$, $\textbf{v}_1=\left(1,1,\ldots,1\right)^\mathsf{T}$ and rewriting $\boldsymbol{\delta}_{\textrm{PC}}$ as a linear combination of the basis of eigenvectors gives
\begin{equation}
\boldsymbol{\delta}'_{\textrm{PC}}(nT)=\boldsymbol{\delta}_{\textrm{PC}}(nT)-\boldsymbol{\delta}_{\infty}=\sum_{i=2}^{N}\beta_{i}a_{i}^n\textbf{v}_i.
\label{anal1}
\end{equation}
Then, since $a_2$ is the second largest eigenvalue, taking the infinity norm in (\ref{anal1}) gives
\begin{align}
d_{\textrm{PC}}(nT)&=\max_{j}\left|\left(\sum_{i=2}^{N}\beta_{i}a_{i}^n \textbf{v}_i\right)_{j}\right| \nonumber \\
&=\left|\beta_2a_2^n\right| \max_{j}\left|\left(\textbf{v}_2+\sum_{i=3}^{N}\frac{\beta_{i}}{\beta_2}\left(\frac{a_{i}}{a_{2}}\right)^n \textbf{v}_i\right)_{j}\right|\nonumber \\
&\sim\left|\beta_2\right|\left|a_2\right|^n \max_{j} \left| \textbf{v}_{2,j}\right|
\label{nonexp}
\end{align}	
where $\sim$ means `is asymptotically equal to (as $n\to\infty$)'.\\
Taking the logarithm gives
\be
\log\left(d_{\textrm{PC}}(nT)\right)\sim n\log\left|a_2\right|
\label{anal2}
\ee
On the other hand $d_{\textrm{PC}}(n)$ asymptotically defines the synchronization time by
\be
d_{\textrm{PC}}(nT)\sim\exp\left(-\frac{nT}{\tau_\textrm{PC}}\right)
\ee
which after taking the logarithm gives
\be
\log(d_{\textrm{PC}}(n))\sim -\frac{nT}{\tau_\textrm{PC}}.
\label{anal3}
\ee
Comparing (\ref{anal2}) with (\ref{anal3}) leads to
\be
\tau_\textrm{PC}=-\frac{T}{\log\left|a_2\right|}.
\label{eq:pcsynctime}
\ee
We numerically find the second largest eigenvalue of the matrix $A$ and use this to calculate the analytical synchronization time (\ref{eq:pcsynctime}). To minimize the influence of specific rewired networks and perturbations we average over $100$ realizations to obtain the average synchronization time $\left\langle \tau_\textrm{PC}\right\rangle$. As for the Kuramoto system, the prediction of synchronization times based on the eigenvalues
of the matrix $A$ well agrees with those obtained from direct numerical simulation (Fig.~\ref{fig:syncscheme}).

\section{The synchronization times for several network ensembles}\label{ensembles}

\begin{figure*}
\includegraphics[width=1.0\textwidth]{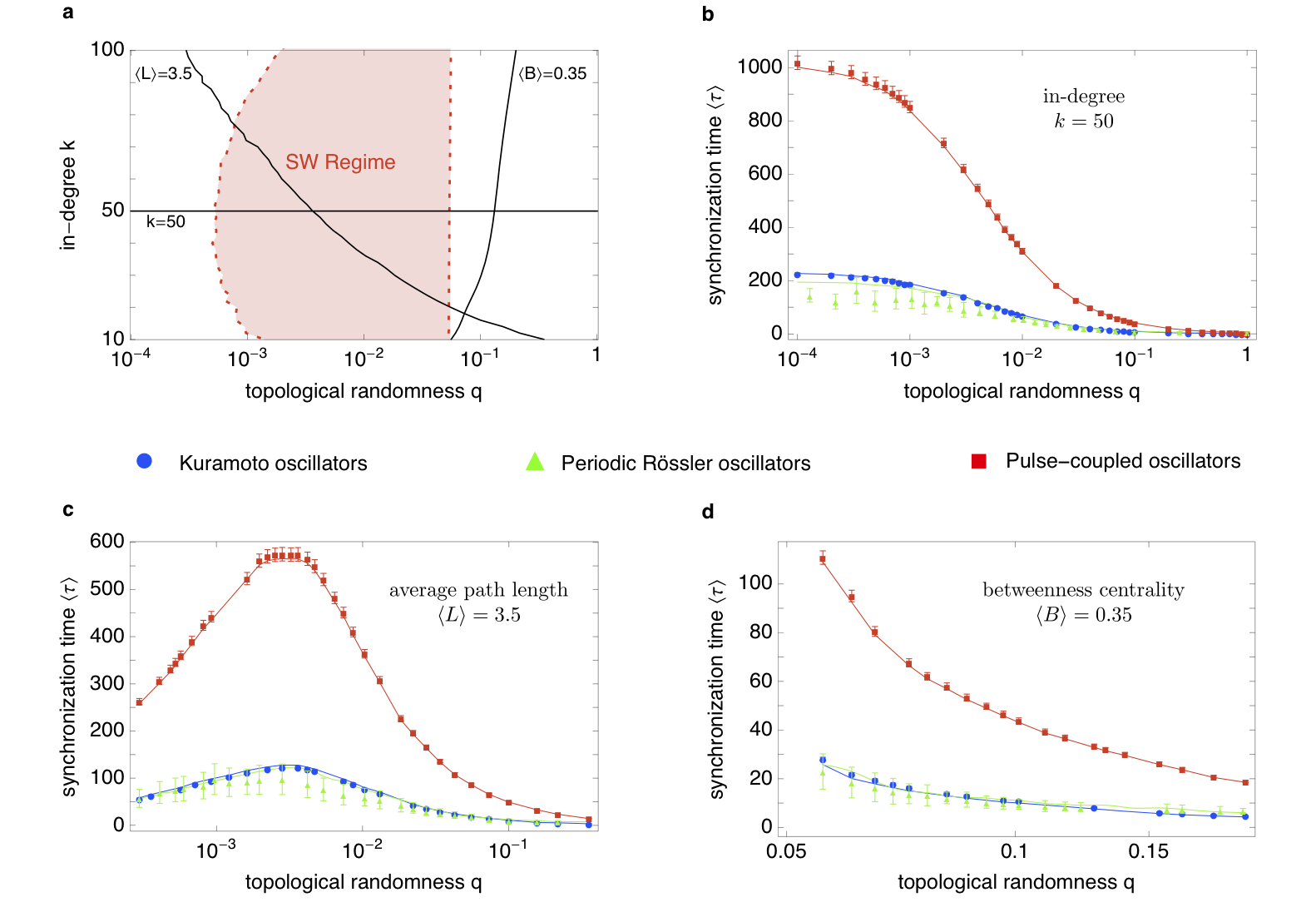}
\caption{(Color online) Comparing the synchronization times of different network ensembles, the small-world regime does not appear to be special.
\textbf{a}: solid lines indicate network ensembles with fixed in-degree $k = 50$, with fixed average path length $\langle L\rangle=3.5$ and with fixed betweenness centrality $\langle B\rangle=0.35$. The small-world regime (\ref{eq:LCdefinitions}) is located between the red
dashed lines. \textbf{b} -- \textbf{d}: analytical predictions based on equations (\ref{eq:kuradistance}), (\ref{eq:roesslerdist}) and (\ref{eq:PCdistance}) (solid lines) and simulation results (symbols with standard deviations) for synchronization times of Kuramoto (blue, circle), R\"ossler (green, triangle) and pulse-coupled oscillators (red, square) for the sketched ensembles. $100$ realizations were carried out in order to average over networks (and in simulations additionally over perturbations).  \label{fig:syncscheme}}
\end{figure*}

Here we study the dependence of the synchronization time on the topological randomness $q$ and on the in-degree $k$. Following the original approach of Watts and Strogatz we first examine ensembles with a fixed in-degree $k$ and then generalize to other generic network network ensembles. 
The results for different dynamics and different ensembles $k(q)$ are depicted in Fig.~\ref{fig:syncscheme}. The synchronization times are studied with the emphasis on their behaviour in the small-world regime which is defined by Eq. (\ref{eq:LCdefinitions}) and highlighted in Fig.~\ref{fig:syncscheme} as the shaded region.

Analytical predictions for the synchronization times in (\ref{eq:kurasynctime}), (\ref{eq:rsynctime}) and (\ref{eq:pcsynctime}) are averaged over $100$  network realizations and depicted in Fig.~\ref{fig:syncscheme} as solid lines. 
Simulation results for the synchronization times are averaged over $100$ realizations of networks and perturbations.

For the Kuramoto and pulse-coupled oscillators, determining the eigenvalues of the stability matrices of networks yields synchronization time estimates that well agree with those found from direct numerical simulations.

Only for the R\"ossler oscillators the synchronization times obtained from the numerical measurement of the decaying maximal distances (\ref{eq:distancedecay}) show small but systematic deviations from the analytically predicted ones. These deviations may be due to inaccuracies in determining the Euclidean distances that oscillate (see the decaying $x$-coordinates only in Fig.~\ref{fig:alldeltas}, \textbf{b}).

\subsection{Networks with fixed in-degree}\label{fixedk}

This subsection is dedicated to the dependence of average synchronization time $\left\langle \tau\left(q,k\right)\right\rangle$ on the topological randomness $q$ for standard Watts-Strogatz ensembles of networks with fixed in-degree $k$. 

We see in Fig.~\ref{fig:alldeltas} and Fig.~\ref{fig:syncscheme}(\textbf{b}) that $\left\langle \tau\left(q,k\right)\right\rangle$ is monotonically decreasing with the topological randomness $q$ and systematically depends on the network topology: Regular ring networks ($q\rightarrow0$) are
typically relatively slow to synchronize. We find that increasing $q$ towards the small-world regime
induces shorter and shorter network synchronization times, with small
worlds synchronizing a few times faster than regular rings. Further increasing the randomness $q$ induces
even much faster synchronization, with fully random networks ($q\rightarrow1$)
synchronizing fastest (two orders of magnitude faster than small worlds in our examples).
Thus in network ensembles with fixed in-degree small worlds just occur intermediately during a monotonic increase of synchronization speed, but are not at all topologically optimal regarding their synchronization time.

One could try to explain this dependence heuristically by the decrease of the average characteristic path length. Indeed, the dependence of $\left\langle L\left(q,k\right)\right\rangle$ on $q$ is also monotonically decreasing in a similar fashion. It is intuitive that as the characteristic path length decreases, oscillators can communicate more efficiently and this leads to faster and more efficient synchronization.

\subsection{Ensembles with fixed average path length or betweenness centrality}\label{fixedL}

We therefore systematically study the synchronization time for generalized Watts-Strogatz ensembles
of networks, specified by a function $k(q)$, where the average path length $\left\langle L\right\rangle $ is fixed while the
degree of randomness $q$ varies. We fix the average characteristic path length $\left\langle L\left(q,k\right)\right\rangle=3.5$ as this gives us a wide range of $q$ values. However, the results below are not sensitive to a change of this specific value, cf. also \cite{Grabow10}. We do not take $k<10$ as the networks are in general no longer strongly connected for larger $q$ values. For each of these in-degrees $k$ a value of the topopolgical randomness $q(k)$ is determined. Note that the standard deviations are larger for smaller $q$ values. This is because we are rewiring a small number of edges here $\left(Nk/q\ \textrm{on average}\right)$ and rewiring one edge more or less may have a strong effect on $L(q,k)$ as it may add a long-range connection where there was not one previously. Note that $k$ decreases in a non-linear fashion as $q$ increases for networks with $\left\langle L\left(q,k\right)\right\rangle=3.5$. When we increase $q$, we decrease the in-degree $k$. Thus, it might be expected that the amount of coupling each oscillator receives also decreases. This would affect the synchronization time \cite{Timme04}. We remove this factor by keeping the input each oscillator receives fixed (\ref{eq:couplingnorm}) as $q$ varies. By doing this, we have reduced the effect of changing the in-degree $k$ on the synchronization time.
Surprisingly, the synchronization time of network ensembles with fixed average characteristic path length non-monotonically depends on the topological randomness (Fig.~\ref{fig:syncscheme}, \textbf{c}). In particular, networks with intermediate randomness in the small-world
regime synchronize slowest. 

Since faster synchronization times are apparently not only related to the decrease of the average path length, we investigated the dependence on other topological observables which have been suggested to control whether or not a network actually synchronizes \cite{Nishikawa:2003p656,Lee06,Goh06,Motter:2005p76,Arenas06,Kurths2006}. 
Representatively, an ensemble with fixed betweenness centrality\footnote{$B(k,q)$ of node $i$  (see e.g. \cite{Freeman77}) measures the number of directed shortest paths that pass through this node, averaged over all nodes $i$.} is shown in ((Fig.~\ref{fig:syncscheme}, \textbf{d})).
Ensembles with fixed clustering coefficient show a similar dependence, but also cover only a small range of $q$ values.

\subsection{Synchronization time of generic network ensembles}\label{nonlin}

\begin{figure}	
\begin{centering}
\includegraphics[width=80mm]{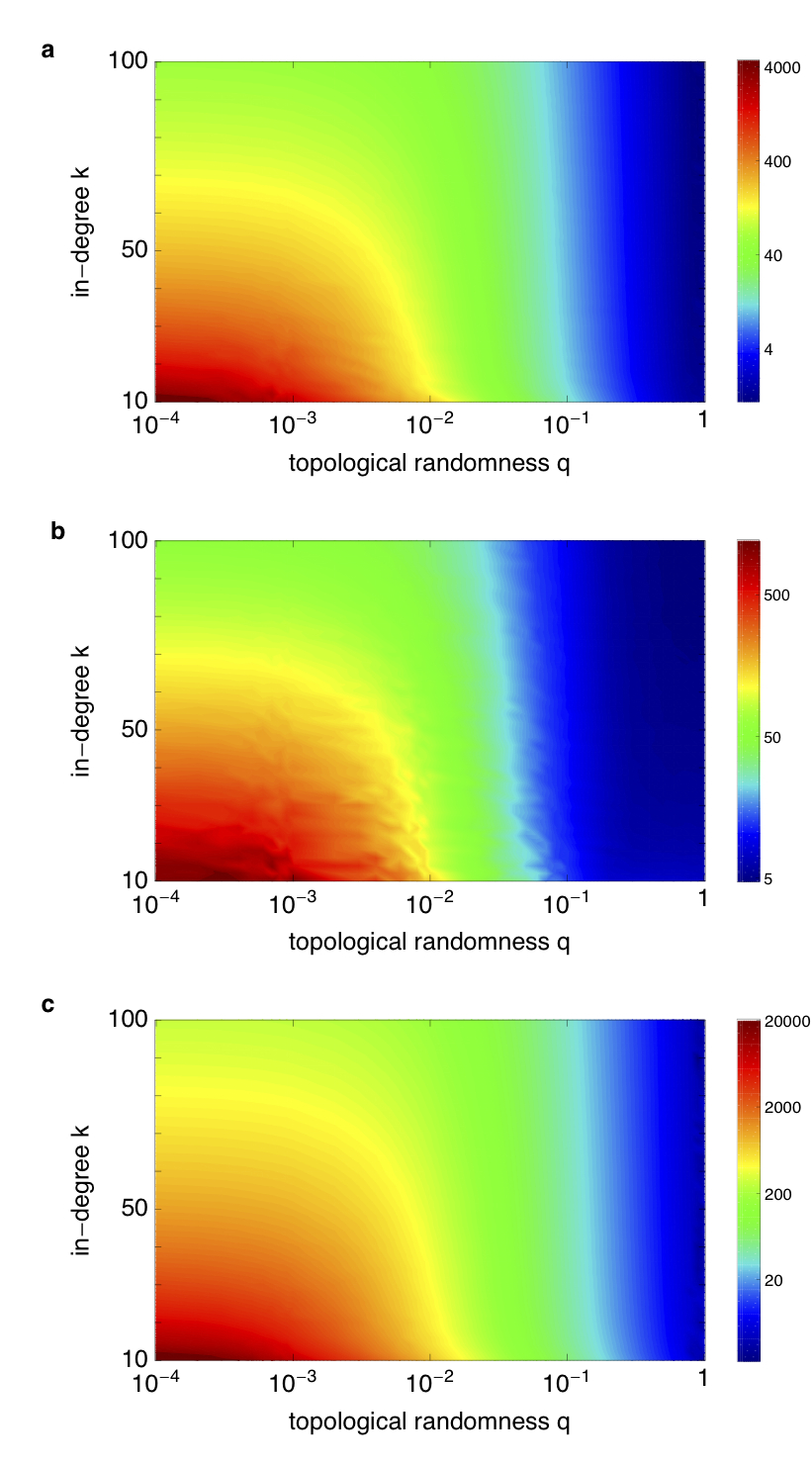} 
\par\end{centering}

\caption{(Color online)
Universal nonlinear dependence of synchronization time on in-degree $k$ and
topological randomness $q$. No generic ensemble $k(q)$ exhibits
fastest synchronization in the small-world regime. \textbf{a}: Kuramoto oscillators (modified from \cite{Grabow10}), \textbf{b}: R\"ossler oscillators, \textbf{c}: pulse-coupled oscillators; logarithmic color scales. Parameters chosen as in Fig.~\ref{fig:alldeltas}. Synchronization times are obtained from equations (\ref{eq:kurasynctime}), (\ref{eq:rsynctime}) and (\ref{eq:pcsynctime}).
\label{fig:syncall} }
\end{figure}

How does synchronization speed vary with randomness for more general ensembles $k(q)$?
A systematic study of the synchronization time as a function of both
in-degree $k$ and randomness $q$ (Fig.~\ref{fig:syncall}) reveals
an interesting nonlinear dependence. 

Firstly, it confirms that for all networks with fixed in-degree $k$ the synchronization time is monotonic
in the randomness $q$ and the small-world regime at intermediate
randomness is not specifically distinguished. 

Secondly, the two-dimensional
function $\left\langle \tau(q,k) \right\rangle$ implies that ensembles of networks with fixed
path lengths all exhibit a non-monotonic behavior of the synchronization
time, with \emph{slowest} synchronization for intermediate randomness.

Thirdly, considering graph ensembles characterized by any other smooth function $k(q)$, $q\in[0,1]$, shows that this is a general phenomenon and the specific choice of an ensemble $k(q)$ is not essential. 

In fact, as illustrated in Fig.~\ref{fig:syncall}, for any generic network ensemble $k(q)$ (including ensembles with fixed in-degree, fixed path length and fixed betweenness centrality as special choice) the synchronization speed $\left\langle \tau(q,k(q))\right\rangle $ is either intermediate or slowest, but never fastest at intermediate randomness, in particular in the small-world regime.
It is remarkable that this seems to hold universally as the synchronization times are similar for Kuramoto oscillators (Fig.~\ref{fig:syncall}, \textbf{a}), periodic (Fig.~\ref{fig:syncall}, \textbf{b}) and chaotic \cite{Grabow10} R\"ossler oscillators and pulse-coupled oscillators (Fig.~\ref{fig:syncall}, \textbf{c}).

\subsection{Relation between Kuramoto and pulse-coupled oscillators}\label{kurapulse}

Comparing the synchronization times for Kuramoto and pulse-coupled oscillators in Fig.~\ref{fig:syncall} both show a striking similarity: Could the synchronization times be mapped on each other?

Therefore let us investigate how perturbations $\boldsymbol{\delta}(t)$ evolve in both systems:
For the Kuramoto oscillators we have
\bel{eq:kuraperturb}
\boldsymbol{\delta}_{\textrm{K}}(t)=e^{\Real \lambda_{2}t}\boldsymbol{\delta}(0)
\ee
while perturbations in the pulse-coupled system propagate like
\bel{eq:PCperturb}
\boldsymbol{\delta}_{\textrm{PC}}(nT)=\left|a_{2}\right|^{n}\boldsymbol{\delta}(0) .
\ee
Obtaining similar synchronization times for both dynamics demands these perturbations to evolve in the same way.
Setting $nT:=t$ the crucial eigenvalues should satisfy
\be
\Real \lambda_{2}= \frac{\log \left|a_{2}\right|}{T}.
\ee
Comparing the structure of the two relevant matrices $A$ (\ref{eq:PCmatrixelements}) and $\Lambda$ (\ref{eq:laplacian}), we obtain the following relation for the respective eigenvalues $a_{2}$ and $\lambda_{2}$:
\be
\lambda_{2}=\frac{c_{\textrm{K}}}{c_{\textrm{PC}}}a_{2}-1
\ee
where the quotient $c_{\textrm{K}}/c_{\textrm{PC}}$ depends on the system parameters
\begin{align}
c_{\textrm{K}}=&\sigma_{\textrm{K}} \\
c_{\textrm{PC}}=&\frac{\gamma \sigma_{\textrm{PC}}}{I e^{- \gamma \Delta} +\gamma \sigma_{\textrm{PC}}} .
\end{align}
Note that $\sigma_{\textrm{K}}>0$ while $\sigma_{\textrm{PC}}<0$.
Setting the parameters in the way that this quotient and the period $T$ ($T$ is close to one anyhow) are both equal to one, we get
\begin{equation}
1+\log \left|a_{2}\right| = \Real a_{2} \quad \text{provided} \ \left|a_{2}\right| \approx 1 \, , \, \Real a_{2} \approx 1.
\end{equation}
This means that synchronization times obtained for networks of pulse-coupled oscillators and for the same structured networks of Kuramoto oscillators are equal if the second largest eigenvalue of matrix $A$ satisfies $\Real a_{2}\approx 1 $ and $\left|a_{2}\right| \approx 1 $, in the sense that $\left|\log \left|a_{2}\right|\right|\ll1$. This is in general the case for large networks, for which the stochastic matrix $A$ has $N$ eigenvalues with real parts in $[-1,1]$.

\section{Real-world networks}\label{real}

\begin{figure*}
\begin{centering}
\includegraphics[width=1.0\textwidth]{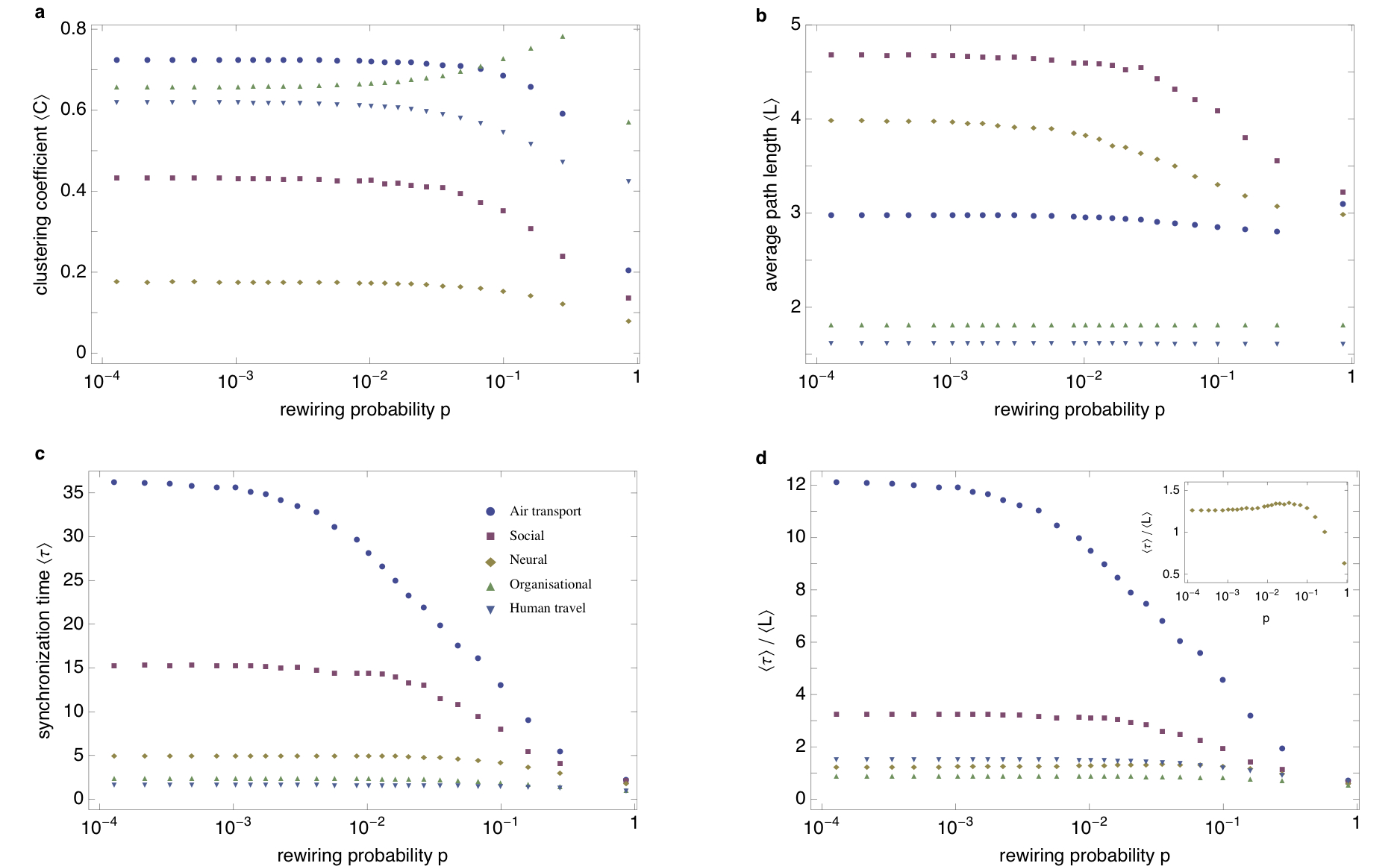} 
\par\end{centering}

\caption{(Color online)
Real-world networks consistently synchronize several times slower than their randomized counterparts (air transport network: the US airports with the largest amount of traffic \cite{colizza}, social network: inmates in prison \cite{prison}, neural network: \textit{C. Elegans} \cite{celegans}, organisational network: research team in a manufacturing company \cite{cross}, human travel network: based on the trajectories of dollar bills \cite{brockmann06}). \textbf{a}: clustering coefficients $ \left\langle C  \right\rangle$. \textbf{b}: average path lengths $\left\langle L  \right\rangle$. \textbf{c}: synchronization times $\left\langle \tau \right\rangle$ \textbf{d}: synchronization times $\left\langle \tau \right\rangle$ relative to the average path length $\left\langle L  \right\rangle$. Inset: the synchronization times for \textit{C. Elegans} divided by the average path length $ \left\langle L  \right\rangle$ show slightly non-monotonic behaviour.  All quantities are averaged over $100$ network realizations.
\label{realnetworks}}
\end{figure*}

For Watts-Strogatz small-world model networks we have found that the synchronization speed is either intermediate or slowest, but never fastest in the small-world regime. Moreover, keeping the in-degree fixed, the model networks synchronize the faster the more random they are. To support that this monotone relation also holds in much more generality we considered various real-world networks: an air transport network \cite{colizza}, a social network \cite{prison}, a neural network \cite{celegans}, an organisational network \cite{cross} and a human travel network \cite{brockmann06}. 

Randomizing is performed by rewiring the existing connections with rewiring probability $p \in [0,1]$. Note that the rewiring probability $p$ and the topological randomness $q$ are two different quantities. Here we start with the original real-world network ($p=0$), which may be in the small-world regime already. The rewiring process is performed as explained in Section \ref{top}: only the outgoing edges of the directed links are rewired, which means that each node's in-degree stays constant during the rewiring. Having considered networks with the same in-degrees for all nodes so far, these in-degrees may vary in real-world networks: But due to the rewiring routine the network's initial in-degree distribution is kept constant during rewiring. Thus it is not uncommon that a network may split from only one strongly-connected component (SCC) in the beginning to several ones \cite{Timme:2006p292}. If splitting occurs the rewiring is repeated until the resulting network consists of one SCC only again.
All measured quantities are averaged over $100$ network realizations.

The studied real-world networks show small-world behaviour: gradually randomizing these networks, their clustering coefficients and their average path lengths monotonically decrease (Fig.~\ref{realnetworks}, \textbf{a} and \textbf{b}). Only the clustering coefficient of the human travel network partly shows a non-monotonic behaviour. This network is extremely large (464670 nodes) in comparison to the other networks ($\approx 100$ -- $500$ nodes).

We found that all real-world networks consistently synchronized several times slower than their randomized counterparts (Fig.~\ref{realnetworks}, \textbf{c}). For all networks, the synchronization times monotonically increase with increasing random
rewiring. For all but the neural circuit of \textit{C. Elegans} \cite{celegans}, this
holds even for ensembles with fixed average path length; for the latter we
observed non-monotonic dependence with slowest synchronization for
intermediate randomness (Fig.~\ref{realnetworks}, \textbf{d}).

\section{Conclusion}\label{conc}

We investigated the effect of topology on the speed of synchronization of various oscillator types, intrinsic dynamics and coupling schemes: phase oscillators coupled via phase differences, higher-dimensional periodic systems coupled diffusively as well as neural circuits with inhibitory delayed pulse-coupling and consistently found qualitatively the same results. 
We derived analytical predictions for the asymptotic synchronization times, including an extension of the master stability function to determine how fast the system actually synchronizes.
We compared the synchronization speed for different network ensembles:

Firstly, we found that for networks of fixed in-degree $k$, the average synchronization time $\left\langle \tau\right\rangle$ is monotonically decreasing with the topological randomness $q$. Comparing different fixed-$k$ ensembles small-world networks always synchronize quicker than regular networks.

Secondly, the intuitive idea that this is due to the decrease in average characteristic path length $\left\langle L(q,k)\right\rangle$ does not provide a complete explanation: Instead of fixing the in-degree, we fixed the average characteristic path length. For such ensembles networks in the small-world regime synchronize slower than regular networks and the synchronization speed non-monotonically depends on the topological randomness $q$. 
The in-degree $k$ is monotonically decreasing with $q$ and so does the clustering coefficient $\left\langle C(q,k)\right\rangle$. So neither of these topological properties alone gives an obvious explanation and the phenomenon results from an interplay between several network properties. For example, the faster synchronization of regular networks than of small-world networks may be due to the in-degree $k$ being large. This is not because the total coupling strength $J_{ij}=k\sigma$ is high, as we kept this fixed for all $(k,q)$-pairs, but may be because the oscillators receive the coupling effect from a large number of oscillators. However, we also see fast synchronization for random networks where $k$ is small and so the same total coupling amount is received from far fewer interacting oscillators. So the explanation for the non-monotonic dependence is non-trivial. 

Thirdly, we investigated the dependence on other topological observables apart from small-world properties: network ensembles with fixed betweenness centrality have been displayed as an example, but yet a simple explanation for the nonlinear dependence is missing.

Further, we studied the full nonlinear dependence of the synchronization time on the in-degree $k$ and the topological randomness $q$ for generic network ensembles. We found that fastest synchronization is essentially impossible in the small-world regime, except for highly artificial ensembles. This statement holds for all observed dynamics. In particular, the synchronization times for the Kuramoto and pulse-coupled oscillators are strikingly similar. 

It would be interesting to extend the analysis started in Section \ref{kurapulse} to find out under which conditions the synchronization times for Kuramoto oscillators could be approximated by or even analytically mapped on the times for the pulse-coupled oscillators. Additionally an understanding or even an analytical description of the curves of same synchronization times in Fig.~\ref{fig:syncall} is extremely helpful for finding further relations between the topology and the dynamics of complex networks. In particular, it is an interesting question to understand the behaviour of the second largest eigenvalue of the Laplacian as a function of $q$ for fixed $k$ and $N$. First results indicate that this might follow a simple power law with exponent close to $1$ for values of $q >1/N$.

In this article we focused on systems with fixed size $N$. In the Watts-Strogatz ensemble the scaling of quantities such as the average path length $L$ or the clustering coefficient $C$ with the system size depends heavily on $q$ \cite{WattsStrogatz98}, e.g. $L \sim N$ for $q=0$ and $L \sim \log N$ for $q=1$. Therefore, it would be an interesting question to study how the results illustrated in Fig.~\ref{fig:syncall} change with $N$ and what would be the appropriate definition of the small-world regime and other generalized ensembles with given structural features as a function of the system size.

We found that small worlds in general never synchronize fastest. Specifically, in networks with fixed average
path length, synchrony is consistently fast for regular rings, fastest for completely
random networks, and slowest in the intermediate small-world regime (Fig. \ref{fig:syncscheme}).
It is an astonishing result that this holds across various oscillator types, intrinsic dynamics and coupling schemes: phase oscillators coupled via phase differences, higher-dimensional periodic systems coupled diffusively as well as neural circuits with inhibitory delayed pulse-coupling. 

In particular, small-world topologies are not at all special and may synchronize orders of magnitude slower than completely random networks. So generically the small-world regime either exhibits slowest synchronization or just exhibits no extremal properties regarding synchronization times. 

Given the variety of the investigated dynamical systems our results indicate that this is a universal phenomenon.

Our investigations of real-world networks support this view. Although the considered networks may be in the small-world regime already, rewiring still strongly increases the synchronization speed, even for ensembles with fixed average path length. It remains an open question why rewiring typically implies faster synchronization.

We thank C. Kirst and D. Brockmann for providing data of real-world networks.  
Supported by the BMBF under grant no. 01GQ1005B, by a grant of the
Max Planck Society to MT and by EPSRC grant no. EP/E501311/1 to SG. SG is also grateful for the hospitality of the Hausdorff Research Institute for Mathematics in Bonn.

\end{document}